\documentclass[10pt,a4paper]{article}
\usepackage{jheppub_kim}
\usepackage{pdflscape}
\usepackage{amsmath}
\usepackage{amssymb}
\usepackage{dcolumn}
\usepackage{bm}
\usepackage{color}
\usepackage{amsmath}
\usepackage{breqn}
\usepackage{graphicx}
\usepackage{epsfig}
\usepackage{caption,subcaption}
\usepackage{amsfonts}
\usepackage{graphicx}
\usepackage{parskip}
\usepackage{dcolumn}
\UseRawInputEncoding

\begin{document}

\title{Reconstruction of $f(Q)$ Gravity from Geodesic Congruence Dynamics: Raychaudhuri Equation}

\author[a]{Tuhina Ghorui,}
\author[b]{Prabir Rudra,}
\author[c]{Farook Rahaman}

\affiliation[a] {Department of Mathematics, Jadavpur University, Kolkata-700 032, India.}

\affiliation[b] {Department of Mathematics, Asutosh College,
Kolkata-700 026, India.}

\affiliation[c] {Department of Mathematics, Jadavpur University, Kolkata-700 032, India.}

\emailAdd{tuhinaghorui.math@gmail.com}
\emailAdd{prudra.math@gmail.com}
\emailAdd{farookrahaman@gmail.com}

\abstract{In this study, we use the Raychaudhuri equation as the essential geometrical tool to reconstruct the functional form of $f(Q)$ gravity within the context of symmetric teleparallel gravity. The expansion scalar and the congruence dynamics recorded in the Raychaudhuri equation are used to relate the non-metricity scalar $Q$ with the kinematical evolution of the Universe in order to establish the reconstruction method. We obtain explicit forms of $f(Q)$ that can explain the observed accelerated expansion of the Universe without the need for an additional dark energy component by taking into account appropriate cosmological assumptions. Moreover, the reconstructed model can also be extended to the infinite past to model the early universe for certain scenarios. In an FLRW spacetime we have considered a constant deceleration parameter and a constant jerk parameter to facilitate our reconstruction analysis. We further discuss the, general relativity limit, Friedmann consistency and background physical viability of the reconstructed models. The analysis indicates that Raychaudhuri equation, interpreted as an equation governing the focusing/defocusing of cosmological geodesic congruences, can be used as the central equation in an inverse reconstruction of $f(Q)$ gravity.}

\maketitle
%%%%%%%%%%%%%%%%%%%%%%%%%%
\section{Introduction}
%%%%%%%%%%%%%%%%%%%%%%%%%%
The question of Dark Energy (DE), which is thought to be the cause of the universe's purportedly rapid expansion, has arguably been the most discussed but unsolved cosmological mystery for the past 20 years \cite{r1, r2}. The observationally necessary value of a non-zero cosmological constant seems to be excessively tiny in comparison to the theoretically anticipated value, even though it can match the observational data \cite{r3}. With a little fine-tuning, quintessence models, like a scalar field with a potential, also do exceptionally well in describing the cosmological data, but there are very few strongly motivated scalar field models that have theoretical particle physics backing. We refer to \cite{r4} for a comprehensive review on DE.

Modifying the theory of gravity instead of introducing exotic matter is a parallel strategy for resolving accelerated expansion. A non-minimally coupled scalar field theory \cite{r5, r6} or a $f(R)$ theory of gravity \cite{r7, r8, r9} are two examples of such approaches. The Ricci scalar $R$ in the Einstein-Hilbert action is generalized to $f(R)$, an analytic function of $R$, in an $f(R)$ gravity model. Higher powers of $R$ in the Einstein-Hilbert action have been shown to cause inflation, or a rapid expansion, in the very early phases \cite{r10}. Therefore, it is a clear way to determine if negative powers of $R$ in the action can result in a late-time acceleration. Every form of $f(R)$ results in a new theory of gravity, hence it is crucial to evaluate the theory against cosmological observations as well as other needs like the stability of the solutions and local astronomy, like the amount of light bending or perihelion shift. The present literature has a few studies along similar lines, such as \cite{r11, r12, r13}. A comprehensive review on $f(R)$ gravity can be found in \cite{r14}.

The literature has two basic formulations of gravity: the curvature ($R\neq 0$, $\tau=0$) and teleparallel ($R=0$, $\tau\neq 0$) formulations, in which $R$ is the scalar curvature and $\tau$ is the scalar torsion. However, in both of these formulations, the non-metricity $Q$ vanishes. $Q$ is the geometric representation of the length fluctuation of a vector in parallel transport. Nonetheless, there is a unique kind of non-metricity that allows fixed-length vectors. In \cite{new1, new2, new3, new4}, the mathematical underpinnings and physical consequences of this particular kind of nonmetricity have been examined. In a third similar formalism of general relativity (GR), it was thought that a non-vanishing non-metricity $Q$ was the elementary geometrical variable in charge of various gravitational interactions. Symmetric Teleparallel Gravity (STG) \cite{q1} is the name given to this theory. Here, the density of the energy-momentum tensor (EMT) is represented by the Einstein pseudotensor.  It should be mentioned that this tensor eventually becomes a real tensor when we geometrically express it. Following extensive research, STG was extended to $f(Q)$ gravity \cite{q2}, also known as coincident general relativity and non-metric gravity. In \cite{q3,q4}, the cosmological consequences of $f(Q)$ gravity and its empirical constraints were examined. The STG framework has experienced several modifications over the last few decades \cite{q5, q6, q7, q8, q9}. It is anticipated that the non-preservation of the energy-momentum tensor and the introduction of an additional force to the geodesic equation of motion will arise from the non-minimal connection between the geometry and matter sectors. This results in STG generalization, such as the $f(Q,T)$ \cite{q10} model and $f(Q,T^2)$ model \cite{q11}. 

A key component of the geometric theory of gravitation is the Raychaudhuri equation (RE), which establishes a deep connection between spacetime geometry and the dynamical behavior of matter. The equation, which was first developed by Raychaudhuri \cite{akr1} in the early 1950s, controls the evolution of congruences of timelike or null curves and is fundamental to modern general relativity, especially when it comes to the development of singularity theorems \cite{sing1, sing2, sing3, sing4}. The RE essentially explains how adjacent worldlines in a congruence focus or defocus as they move through spacetime. The kinematical characteristics of the flow can be broken down into expansion, shear, and vorticity, given a congruence defined by a tangent vector field. Shear encodes shape distortions, vorticity quantifies local rotation, and the expansion scalar estimates the fractional rate of change of an infinitesimal volume element conveyed along the congruence. In terms of these parameters and the spacetime curvature, the RE gives a precise, purely geometric evolution law for the expansion scalar. The equation's independence from the gravitational field equations is one of its most notable characteristics. It can be applied to any theory of gravity because it is obtained only from differential geometry and the formulation of the Ricci tensor. The RE, when paired with the Einstein field equations and appropriate energy conditions, shows that gravity is intrinsically attractive: positive energy density generally results in the focusing of geodesic congruences. Caustics and, more importantly, spacetime singularities are formed as a result of this focusing action. By connecting the growth of cosmic fluxes to matter content and curvature, RE provides a profound understanding of the universe's evolution.

There are several models that do account for the universe's accelerated expansion, but very few of them satisfy an urgent demand required by other fields of physics. A reconstruction of models from observational data becomes an excellent choice in the absence of a theoretical model that is a clear winner as dark energy. The goal is to determine the necessary matter distribution based on the universe's evolutionary history \cite{r15}. In the present work we attempt to reconstruct $f(Q)$ gravity models from the perspective of Raychaudhuri equation \cite{w1, w2}. The specific purpose of the present work is to investigate whether the Raychaudhuri equation, interpreted as an equation governing the focusing/defocusing of cosmological geodesic congruences, can be used as the central equation in an inverse reconstruction of $f(Q)$ gravity. This is a new approach (very rarely found in the literature) and herein lies the originality and novelty of this work. We consider prescribed kinematical expansion histories and investigate the inverse problem of determining the $f(Q)$ functions that are compatible with these histories through the Raychaudhuri equation. The resulting reconstructed functions are subsequently subjected to the independent Friedmann constraint. This method seems to be a type of reverse procedure which can found in literature \cite{rec1, rec2, rec3}. For the reconstruction, we employ the Raychaudhuri equation, appropriately adjusted for $f(Q)$ gravity. Using the provided expansion history, we create the kinematical quantities and write the metric components in terms of the non-metricity scalar $Q$. Then we integrate the Raychaudhuri equation to find the expression for $f(Q)$.

The Raychaudhuri equation can function just as well in modified gravity theories because it just requires Riemannian geometry at first. Therefore, even without exactly accounting for the metric, this equation can be quite helpful in drawing some broad conclusions about the model. We directly use the Raychaudhuri equation, as a tool, to determine $f(Q)$ gravity models for different scenarios. The Raychaudhuri equation itself is a geometrical identity and therefore remains unchanged in modified gravity. The modification enters through the gravitational field equations used to express the Ricci focusing term in terms of the matter and non-metricity sectors. The motivation for this approach is that the Raychaudhuri equation directly connects the geometrical evolution of congruences with the cosmological expansion. Thus, the conceptual contribution is not the general idea of reconstruction, but the use of geodesic-congruence dynamics as the starting point of the inverse reconstruction procedure in symmetric teleparallel gravity. The paper is organized as follows: In section 2 we present an overview of symmetric teleparallel gravity. In section 3 we discuss Raychaudhuri equation in $f(Q)$ gravity. Section 4 is dedicated to reconstruction of $f(Q)$ gravity models using Raychaudhuri equation. Finally the paper ends with a conclusion in section 5.

%%%%%%%%%%%%%%%%%%%%%%%%%%%%%%%%%%%%%%%%%%%%%%%%%%%%%%%%%
\section{An overview of Symmetric Teleparallel Gravity}
%%%%%%%%%%%%%%%%%%%%%%%%%%%%%%%%%%%%%%%%%%%%%%%%%%%%%%%%%
We start with the action of $f(Q)$ gravity given by
\begin{equation}
S=\int d^{4}x\sqrt{-g}\left(\frac{1}{2}f(Q)+\mathcal{L}_{m}\right)
\end{equation}
where $g \equiv det (g_{\mu\nu})$ is the determinant of the metric tensor $g_{\mu\nu}$, and the Lagrangian of matter is given by $\mathcal L_{m}$. The field equations in $f(Q)$ gravity becomes
\begin{equation}
\frac{2}{\sqrt{-g}} \nabla_{\alpha}\left(f_{Q}\sqrt{-g}P^{\alpha}_{\mu\nu}\right)+\frac{1}{2}f(Q)g_{\mu\nu}+f_{Q}\left(P_{\mu\alpha\beta}Q_{\nu}^{\alpha\beta}-2Q^{\alpha\beta}_{\mu}P_{\alpha\beta\nu}\right)= -T_{\mu\nu}
\end{equation}
where $f_{Q}$ represents $df/dQ$ and the energy momentum tensor, denoted by $T_{\mu\nu}$, is defined as
\begin{equation}
T_{\mu\nu}=-\frac{2}{\sqrt{-g}}\frac{\delta\sqrt{-g}\mathcal{L}_m}{\delta g_{\mu\nu}}
\end{equation}
The field equations can be written in the simplified form
\begin{equation}\label{b1}
f_{Q}G_{\mu\nu}+\frac{1}{2}g_{\mu\nu}\left(Qf_{Q}-f\right)
+2\nabla_{\alpha}P^{\lambda}_{\mu\nu}f_{Q}=T_{\mu\nu}\end{equation}
where $P^{\lambda}_{\mu\nu}$ is the non-metricity superpotential tensor. The field equations of $f(Q)$ gravity can be expressed in terms of the Einstein tensor using an effective energy-momentum tensor $T^{eff}_{\mu\nu}$, which accounts for the curvature contribution as
\begin{equation}
G_{\mu\nu}=\frac{T_{\mu\nu}^{eff}}{f_{Q}}
\end{equation}
where,
\begin{equation}
T_{\mu\nu}^{eff}=T_{\mu\nu}+\frac{1}{2}g_{\mu\nu}\left(f-Qf_{Q}\right)+2\nabla_{\alpha}P^{\alpha}_{\mu\nu}f_{Q}
\end{equation}
At least technically, this equation resembles the Einstein equations; the only difference is the non-minimal coupling indicated by the presence of $f_Q$. The effective gravitational coupling in this model will not be constant. The Einstein field equations in the standard form are given by
\begin{equation}
G_{\mu\nu}=R_{\mu\nu}-\frac{1}{2}Rg_{\mu\nu}
\end{equation}
Taking the trace of this Einstein-like form we get
\begin{equation}
R=-g^{\mu\nu}G_{\mu\nu}
\end{equation}
which gives
\begin{equation}
R =-\frac{1}{f_{Q}}\left[T+2\left(f-Qf_{Q}\right)+2g^{\mu\nu}\nabla_{\alpha}P^{\alpha}_{\mu\nu}f_{Q}\right]
\end{equation}

%%%%%%%%%%%%%%%%%%%%%%%%%%%%%%%%%%%%%%%%%%%%%%%%%%%%%%
\section{Raychaudhuri Equation and $f(Q)$ Gravity}
%%%%%%%%%%%%%%%%%%%%%%%%%%%%%%%%%%%%%%%%%%%%%%%%%%%%%%
For a timelike congruence with velocity vector $u$, the Raychaudhuri equation is given by
\begin{equation}\label{tu1}
\frac{d\theta}{d\tau}=-\frac{1}{3}\theta^{2}+\nabla_{\mu}a^{\mu}-\sigma_{\mu\nu}\sigma^{\mu\nu}+\omega_{\mu\nu}\omega^{\mu\nu}-R_{\mu\nu}u^{\mu}u^{\nu}
\end{equation}
where $\theta=\nabla_{\mu}u^{\mu}$ is the expansion factor, $n^{\mu}=$null vector;

$\sigma^{\mu\nu}=\nabla_{(\nu} u_{\mu)}-\frac{1}{3}\theta+a_{(\gamma}u_{\mu)}$ is the shear tensor;

$\omega_{\mu\nu}=\nabla_{[\nu}u_{\mu]}-a_{[\nu}u_{\mu]}$ is the rotation tensor;

$h_{\mu\nu}$ is the spatial metric;

$a^{\mu}=u^{\mu}\nabla_{\nu}u^{\mu}$ is the acceleration vector.

$R_{\mu\nu}$ is the Ricci tensor and $u_{\mu}$ is a timelike vector.

The final term on the right-hand side of the equation (\ref{tu1})  can be expressed using the field equations (\ref{b1}) for f(Q) theory as

\begin{equation}\label{riccci}
R_{\mu\nu}=\frac{1}{f_{Q}}\left[T_{\mu\nu}-\frac{1}{2}Tg_{\mu\nu}-\frac{1}{2}g_{\mu\nu}\left(f-Qf_{Q}\right)-2\nabla_{\alpha}(f_{Q}P^{\alpha}_{\mu\nu})\right]
\end{equation}

We shall now examine the conventional Friedmann-Lemaitre-Robertson-Walker (FLRW) line element, which is expressed explicitly as
\begin{equation}\label{frr}
ds^{2}=-dt^{2}+a^{2}(t)\left[dr^{2}+r^{2}\left(d\theta^{2}+sin^{2}\theta d\phi^{2}\right)\right]
\end{equation}
where $a(t)$ is the cosmological scale factor. For the FLRW symmetry, the shear $\sigma^{\mu\nu}$, rotation $\omega_{\mu\nu}$ and acceleration $a^\mu$ vanishes and we are left with $\frac{d\theta}{d\tau}=-\frac{1}{3}\theta^{2}-R_{\mu\nu}u^{\mu}u^{\nu}$. We consider matter distribution in the form of perfect fluid given by $T^{\mu\nu}=\left(\rho+p\right)u^{\mu}u^{\nu}+pg^{\mu\nu}$. The FLRW equations for $f(Q)$ gravity are given by
\begin{equation}\label{1frw}
6f_{Q}H^{2}-\frac{1}{2}f=\rho
\end{equation}
and
\begin{equation}\label{rayc}
\left(f_{Q}+2Qf_{QQ}\right)\dot{H}=-\frac{1}{2}\left(\rho+p\right)
\end{equation}
The second equation is the Raychaudhuri equation. For the FLRW spacetime, the nonmetricity $Q$ is given by $Q=6H^2$. From eqn.(\ref{riccci}) we get
\begin{equation}
R_{\mu\nu}u^{\mu}u^{\nu}=\frac{1}{f_{Q}}\left[\rho+\frac{1}{2}T+\frac{1}{2}\left(f-Qf_{Q}\right)-3H\dot{f_{Q}}\right]
\end{equation}

For the FLRW spacetime we can get the acceleration equation by modifying the Raychaudhuri equation as
\begin{equation}
\frac{\ddot{a}}{a}=H^{2}+\dot{H}=-\frac{1}{3}R_{\mu\nu}u^{\mu}u^{\nu}
\end{equation}
which gives
\begin{equation}\label{m1}
\frac{\ddot{a}}{a}=\frac{Q}{6}-\frac{\rho+p}{2\left(f_{Q}+2Qf_{QQ}\right)}
\end{equation}

%%%%%%%%%%%%%%%%%%%%%%%%%%%%%%%%%%%%%%%%%%%%%%%%%%%%%%%%
\section{Reconstruction of $f(Q)$ Gravity models using Raychaudhuri equation}
%%%%%%%%%%%%%%%%%%%%%%%%%%%%%%%%%%%%%%%%%%%%%%%%%%%%%%%%
Since the proposal of $f(Q)$ gravity, various models have been used to study the universe. But none of these models is self-sufficient to explain the evolution of the universe. So there is good motivation to construct new $f(Q)$ models from different mechanisms. 

It is important to clarify the nature of the reconstruction procedure adopted in this work. Instead of performing a forward reconstruction where a specific form of $f(Q)$ is assumed and the cosmic dynamics are then solved, this reconstruction is inverse or kinematical. The current method uses a kinematical condition, such as a constant jerk value or a constant deceleration parameter, to specify the cosmic expansion history. After obtaining the corresponding scale factor from the kinematical relation, the Raychaudhuri equation is utilized to find the $f(Q)$ function compatible with the specified expansion history using the resulting $H(t)$ and non-metricity scalar $Q=6H^2$. Thus, the reconstruction procedure can be schematically represented as $a(t)\longrightarrow H(t)\longrightarrow Q(t)
\longrightarrow {\rm Raychaudhuri\ equation}\longrightarrow f(Q).$ Therefore, the scale factor in this process should not be understood as an independently derived solution to a $f(Q)$ theory provided as a priori. Rather, it depicts the kinematical history that is used to recover the associated gravitational Lagrangian. The distinction is crucial because, although the Raychaudhuri equation supplies the dynamical equation needed for the reconstruction, it does not take the place of the entire set of Friedmann equations on its own. Instead, it is derived from the gravitational field equations. Therefore, to represent a self-consistent cosmological solution, a reconstructed $f(Q)$ must meet the remaining independent Friedmann constraint. So the logic can be presented as:\\\\ Prescribed kinematical history $a(t)$ $\longrightarrow$ $H(t)$, $\dot{H}(t)$, $Q(t)=6H^{2}(t)$ $\longrightarrow$ Raychaudhuri Equation $\longrightarrow$ reconstructed $f(Q)$ $\longrightarrow$ Friedmann consistency condition $\longrightarrow$ admissible reconstructed $f(Q)$ models.

We will use equation (\ref{m1}) to reconstruct $f(Q)$ gravity models for an accelerating universe scenario. Kinematical characteristics such as the jerk parameter $j$ or the deceleration parameter $q$ will determine the mode of acceleration. Here, two scenarios are examined: one in which the cosmos is always accelerating with a constant acceleration value, and the other in which the jerk parameter $j=1$ denotes a model that resembles the $\Lambda$CDM model in conventional general relativity.

%%%%%%%%%%%%%%%%%%%%%%%%%%%%%%%%%%%%%%%%%%%%%%%%%%%%%%%%%%%%
\subsection{A constant deceleration parameter $q$}
%%%%%%%%%%%%%%%%%%%%%%%%%%%%%%%%%%%%%%%%%%%%%%%%%%%%%%%%%%%%%
We know that the Hubble parameter is expressed as a first order time derivative of the cosmological scale factor. It is connected to the expansion of the universe and is a dynamically evolving quantity. the next higher order derivative of the scale factor is used to construct the the deceleration parameter, which determines the nature (accelerating or decelerating) of the cosmological expansion.
The deceleration parameter is given by
\begin{equation}\label{t1}
q=-\frac{a\ddot{a}}{\dot{a}^{2}}
\end{equation}
Here we will consider a constant negative deceleration parameter. Although this ensures an accelerated expansion of the universe, a constant deceleration parameter is not observationally favored. But this is a useful approximation for the accelerated universe, providing a phenomenological description of the universe. This consideration makes the cosmological equations analytically solvable, because it gives closed-form expressions for the scale factor $a(t)$. This is extremely useful for building toy-models, testing theoretical ideas, and comparing with observational data. The above expression for $q$ can be written in the form
\begin{equation}\label{t11}
q=-\frac{1}{H^2}\frac{\ddot{a}}{a}=\frac{1-m}{m}=\mu=constant
\end{equation}
where $m$ is a constant and $H=\frac{\dot{a}}{a}$ is the Hubble parameter. Now from the observations we can put constraints on the values of $m$. We see that to get a negative deceleration parameter we should have $m>1$. Solving eqn.(\ref{t11}) we get the scale factor as
\begin{equation}\label{t2}
a(t)=a_{0}\left(t-t_{0}\right)^{m}
\end{equation}
where $a_{0}$ and $t_{0}$ are integrating constants.
This is a power-law solution which is an acceptable solution to explain the evolution of the universe. It should be mentioned here that the constant in eqn.(\ref{t11}) is considered in an unorthodox way so that the final solution of the scale factor in eqn.(\ref{t2}) is obtained in a convenient power-law form. We start with the Friedmann equations given by
\begin{equation}\label{t4}
H^{2}=\frac{1}{3}\rho_{eff}
\end{equation}
\begin{equation}\label{t18}
\dot{H} + H^{2}=\frac{\ddot{a}}{a}=-\frac{1}{6}\left(\rho_{eff}+3p_{eff}\right)
\end{equation}
Using eqns.(\ref{t2}) and (\ref{t4}) we get,
\begin{equation}\label{t3}
3H^{2}=\rho_{eff}=\frac{3m^{2}}{\left(t-t_{0}\right)^{2}}
\end{equation}

The above equation can be put in the following form using the non-metricity scalar $Q$ 
\begin{equation}
a\left(t\right)=a_{0}\left(\frac{6m^{2}}{Q}\right)^{\frac{m}{2}}
\end{equation}
where $Q=6H^{2}$. The second FRW equation can be written in the form
\begin{equation}\label{t6}
\frac{\ddot{a}}{a}=\frac{m\left(m-1\right)}{\left(t-t_{0}\right)^{2}}
\end{equation}
Using Equations (\ref{t2}) and (\ref{t18}), it can also be written as
\begin{equation}\label{t9}
6\left(\dot{H} + H^{2}\right)=-\left(\rho_{eff}+3p_{eff}\right)= -\frac{6m\left(m-1\right)}{\left(t-t_{0}\right)^{2}}
\end{equation}
This actually gives the effective energy condition. For $m>1$, this energy condition is negative, which is expected for an accelerated universe. From equation (\ref{t3}) and (\ref{t9}), we get
\begin{equation}\label{t8}
p_{eff}=\frac{-3m^{2}+2m}{\left(t-t_{0}\right)^{2}}
\end{equation}
The equation of state (EoS) parameter is given by $w_{eff}=\frac{p_{eff}}{\rho_{eff}}$. Using (\ref{t3}) and (\ref{t8}) we get the EoS parameter as
\begin{equation}
w_{eff}=-1+\frac{2}{3m}
\end{equation}
If $m=\frac{2}{3}$ implies that $w_{eff}=0$ (matter era). Further, if $m=\frac{1}{2}$ implies that $w_{eff}=\frac{1}{3}$ (radiation era). From eqn.(\ref{t11}) we have $m=\frac{1}{1+q}$ and $q=-1$ implies that $m\rightarrow\infty$ and $w_{eff}\rightarrow -1$ (de-Sitter like evolution). Thus all the cosmological eras can be achieved from this ansatz. The non-metricity scalar is obtained as,
\begin{equation}
Q=6H^{2}=6\frac{m^{2}}{\left(t-t_{0}\right)^{2}}
\end{equation}
From (\ref{rayc}) we get
\begin{equation}\label{raycc}
f_{Q}+2Qf_{QQ}=\frac{3m\left(\rho+p\right)}{Q}
\end{equation}
We assume that the energy-momentum tensor associated with the fluid distribution is independently preserved, i.e. that the equation $\dot{\rho}+3H(\rho+p)=0$ is satisfied. This conservation equation leads to the matter density $\rho=\rho_{0}a^{-3}$. Using (\ref{raycc}), for the dust-dominated case $(p=0)$ we get the following differential equation
\begin{equation}\label{bh6}
 f_{Q}+2Qf_{QQ}=C_{m}Q^{\frac{3m}{2}-1}
\end{equation} 
where $C_{m}=\frac{3m\rho_{0}}{a_{0}^{3}}\left(6m^{2}\right)^{-\frac{3m}{2}}$. This is a first-order equation for $f_{Q}$, integrating which, yields the solution for $f(Q)$ as
\begin{equation}\label{recon1}
f(Q)=C_{0}+C_{1}\sqrt{Q}+C_{2}Q^{3m/2}
\end{equation}
where $C_0$, $C_{1}$ and $C_2=\frac{2C_{m}}{3m\left(3m-1\right)}$, $m\neq 1/3$ are constants. For $m=2/3$, $C_{0}=C_{1}=0$, $C_{2}=1$, we get $f(Q)=Q$. This is the GR limit of the reconstructed $f(Q)$ model. But in the viable range $m>1$, we do not have a GR limit for this reconstructed model, which is a limitation.

Now we will impose the first Friedmann equation, which will act as a constraint on the reconstructed model, and help to restrict the constants. Using equation (\ref{recon1}) in the first FLRW equation (\ref{1frw}) we get
\begin{equation}
\left(3m-1\right)C_{2}Q^{3m/2}-C_{0}=\frac{2\rho_{0}}{a_{0}^{3}m^{3m}6^{3m/2}}Q^{3m/2}
\end{equation}
Comparing both sides we get $C_{0}=0, C_{2}=\frac{2\rho_{*}}{6^{3m/2}m^{3m}\left(3m-1\right)}$, where $\rho_{*}=\frac{\rho_{0}}{a_{0}^{3}}$. So the fully consistent solution for the constant deceleration parameter case is
\begin{equation}
f(Q)=C_{1}\sqrt{Q}+\frac{2\rho_{*}}{6^{3m/2}m^{3m}\left(3m-1\right)}Q^{3m/2}
\end{equation}
which can be further written as
\begin{equation}\label{condec}
f(Q)=C_{1}\sqrt{Q}+\frac{2\rho_{0}}{6^{3m/2}m^{3m}\left(3m-1\right)a_{0}^{3}}Q^{3m/2}, ~~~Q>0
\end{equation}
It should be noted that, since $C_{0}$ has been eliminated and $C_{2}$ has been calculated in terms of $m$, an effective reduction in the degree of freedom of the system has been achieved using the Friedmann equation, which is a very important step of this reconstruction scheme. The above reconstructed equation may be written in a simplified form as
\begin{equation}
f(Q)=C_{1}\sqrt{Q}+DQ^{3m/2}
\end{equation}
where $D=\frac{2\rho_{0}}{6^{3m/2}m^{3m}\left(3m-1\right)a_{0}^{3}}$. The $\sqrt{Q}$ arises from the homogeneous solution of the equation. So this term actually disappears from the both FLRW and the Raychaudhury equation and the model satisfies both independent background equations identically. When we try to decode the role of $C_{1}$ in the reconstructed model, we see that the $C_{1}$ actually occurs in the model in a background degenerate mode. Consequently, it cannot be fixed by the background cosmological dynamics alone. This is why it remains as an arbitrary integration constant. That is a useful physical observation.

Now we would like to check the reconstructed model for background viability. In order to do that we will check the derivatives of the model and seek their positive region. For the model to be stable we should have $f'(Q)>0$ and $f''(Q)>0$. Using the reconstructed model given in eqn.(\ref{condec}) we have 
\begin{equation}
f'(Q)=\frac{C_{1}}{2\sqrt{Q}}+\frac{3mD}{2}Q^{\frac{3m}{2}-1}
\end{equation}
and
\begin{equation}
f''(Q)=-\frac{C_{1}}{4Q\sqrt{Q}}+\frac{3m\left(3m-2\right)D}{4}Q^{\frac{3m}{2}-2}
\end{equation}

\begin{figure}[hbt!]
\begin{center}
\includegraphics[height=2.5in]{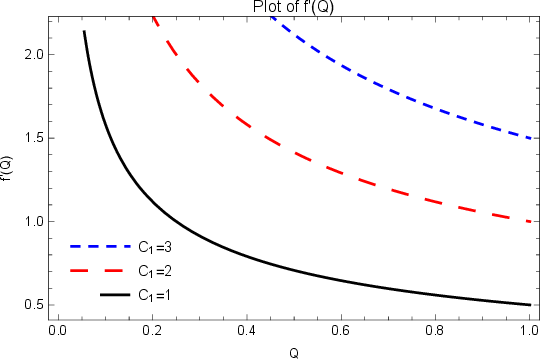}
\caption{Plot of the derivative $f'(Q)$ against $Q$ for different positive values of $C_1$. We have considered $m=1.1, a_{0}=1, \rho_{0}=10^{-27}$.}
\label{fig2}
\end{center}
\end{figure}

\begin{figure}[hbt!]
\begin{center}
\includegraphics[height=2.5in]{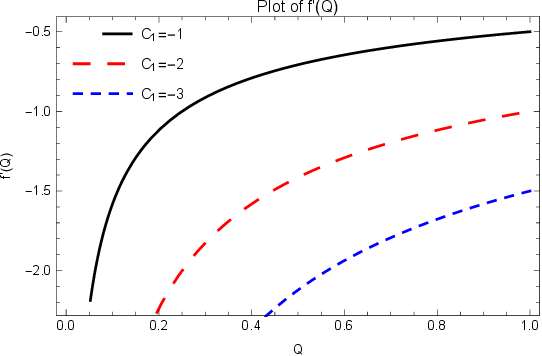}
\caption{Plot of the derivative $f'(Q)$ against $Q$ for different negative values of $C_1$. We have considered $m=1.1, a_{0}=1, \rho_{0}=10^{-27}$.}
\label{fig22}
\end{center}
\end{figure}

In the viable range $m>1$, we have $D>0$. So for $C_{1}=0$, we have $f'(Q)>0$, for $Q>0$. If $C_{1}>0$, then both terms are positive and we have $f'(Q)>0$. We see that $\sqrt{Q}$ term dominates as $Q\to 0$, i.e., $f'(Q)\sim \frac{C_{1}}{2\sqrt{Q}}$. So for $C_{1}<0$, $f'(Q)\to -\infty$, which is undesirable. So if we want to maintain $f'(Q)>0$, throughout the cosmological evolution, a conservative condition should be $C_{1}\geq 0$. This is quite evident in figs.(\ref{fig2}) and (\ref{fig22}) where we have plotted $f'(Q)$ against $Q$ for positive and negative values of $C_1$ respectively.

\begin{figure}[hbt!]
\begin{center}
\includegraphics[height=2.5in]{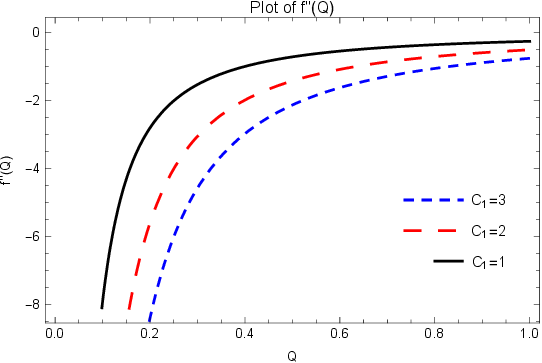}
\caption{Plot of the second derivative $f''(Q)$ against $Q$ for different positive values of $C_1$. We have considered $m=1.1, a_{0}=1, \rho_{0}=10^{-27}$.}
\label{fig3}
\end{center}
\end{figure}

\begin{figure}[hbt!]
\begin{center}
\includegraphics[height=2.5in]{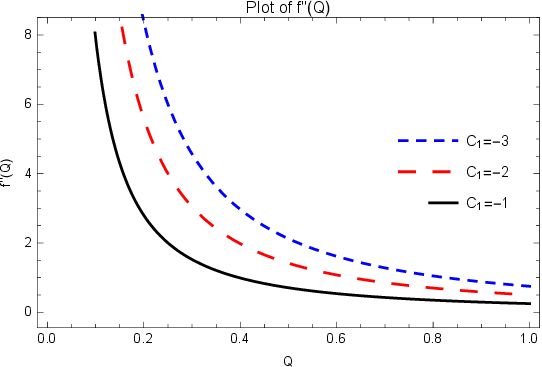}
\caption{Plot of the second derivative $f''(Q)$ against $Q$ for different negative values of $C_1$. We have considered $m=1.1, a_{0}=1, \rho_{0}=10^{-27}$.}
\label{fig33}
\end{center}
\end{figure}

For $C_{1}=0$, $f''(Q)=\frac{3m\left(3m-2\right)D}{4}Q^{\frac{3m}{2}-2}$. For $m>1$, $D>0$ we have $f''(Q)>0$. In figs.(\ref{fig3}) and (\ref{fig33}) we have plotted $f''(Q)$ against $Q$ for positive and negative values of $C_{1}$ respectively. We see that for $C_{1}>0$, we have $f''(Q)<0$ and vice versa. So combining all the above results we see that for only $C_{1}=0$, we simultaneously have $f'(Q)>0$ and $f''(Q)>0$. So we are left with a power law type $f(Q)$ model, i.e., $f(Q)=DQ^{3m/2}$, which satisfies all the background viability conditions. These conditions characterize background viability but do not constitute a complete perturbative ghost analysis. This constant deceleration case actually acts as an analytical benchmark for our inverse reconstruction scheme.

Now we have to understand what the above discussion actually means physically (beyond the mere mathematical intricacies). In $f(Q)$ gravity, the scalar $Q$ encodes non-metricity, i.e. how much the connection fails to preserve the metric ($\nabla_{\alpha}g_{\mu\nu}\neq 0$). So higher or lower $Q$ actually indicates how strongly lengths and angles change under parallel transport. Small $Q$ indicates that the geometry is close to the Riemannian GR-like geometry. Parallel transport almost preserves lengths and angles and the connection behaves nearly like Levi-Civita. Moreover, for this type of spacetime the transported rulers and clocks behave normally. Large $Q$ indicates that there is a strong deviation from metric compatibility. Parallel transport results in significant change in lengths and angles. The affine structure dominates over metric structure and the geometry becomes strongly non-Riemannian. In such a case, the notion of distance becomes path-dependent in a stronger sense. In the FLRW universe, the implication is much more concrete. Since $Q=6H^2$, a large value of $Q$, i.e. $Q>>1$ indicates a large value of $H$. This indicates the early universe scenario dominated by radiation and an inflationary rapid expansion resulting from the strong non-metricity effects. For a low $Q$, i.e. $Q\rightarrow 0$, we have $H\rightarrow 0$, i.e., small $H$. This indicates the late-time universe, where the geometry approaches GR-like behavior.

%%%%%%%%%%%%%%%%%%%%%%%%%%%%%%%%%%%%%%%%%%
\subsection{A constant Jerk Parameter}
%%%%%%%%%%%%%%%%%%%%%%%%%%%%%%%%%%%%%%%%%%
Just like the previous section, where we have considered a constant deceleration parameter, here we will consider a constant jerk parameter and try to reconstruct $f(Q)$ models from it. In cosmology, the jerk parameter $j$ measures the third derivative of the scale factor with respect to cosmic time. The corresponding expression is given by
\begin{equation}
j=\frac{1}{H^{3}}\frac{\dddot{a}}{a}
\end{equation}
Here we consider $j=1$, and find the scale factor solution as
\begin{equation}\label{bi2}
a(t)=\left(Ae^{\lambda{t}}+Be^{-\lambda{t}}\right)^{\frac{2}{3}}
\end{equation}
where $A,B$ and $\lambda$ are the integration constants. It is useful to define $\tau=t-t_{0},~~~ x=\lambda \tau$.

The expression for the scale factor given in eqn.(\ref{bi2}) can be used to compute the effective equation of the state parameter, which is provided by
\begin{equation}
\omega_{eff}=\frac{p_{eff}}{\rho_{eff}}=-\left(\frac{Ae^{\lambda{t}}+Be^{-\lambda{t}}}{Ae^{\lambda{t}}-Be^{-\lambda{t}}}\right)^{2}
\end{equation}
There are two physically different branches. We consider them separately.

\subsection{Type-1 evolution of scale factor}
If $AB<0$, the scale factor can be written as
\begin{equation}
a(t)=a_{0}[\sinh{\left(\lambda\left(t-t_{0}\right)\right)}]^{\frac{2}{3}}
\end{equation}
where $\sinh$ is the sine hyperbolic function. We consider $\tau=t-t_{0}>0$, so that $a>0$. The effective equation of state parameter is given by,
\begin{equation}
\omega_{eff}= \frac{p_{eff}}{\rho_{eff}}=-\tanh{[\lambda\left(t-t_{0}\right)]^{2}}
\end{equation}
where $\tanh$ is the tan hyperbolic function. From this expression we see that $\omega_{eff}\rightarrow -1$ as $t\rightarrow \infty$, and $\omega_{eff}\rightarrow 0$ as $t\rightarrow t_{0}$. This shows that initially we have a matter-dominated era followed by an accelerated expansion. This is perfectly in accordance with the $\Lambda$CDM cosmology. 

For this evolution the Hubble parameter is given by $H=\frac{2\lambda}{3}\coth{x}$. Consequently the non-metricity scalar is given by $Q=Q_{0}\coth^{2}{x}$, where $Q_{0}=8\lambda^{2}/3$. It is straightforward to see that $Q>Q_{0}=8\lambda^{2}/3$. Using eqn.(\ref{rayc}) we get the reconstruction equation for this type of evolution as
\begin{equation}
2Qf_{QQ}+f_{Q}=C_{3}
\end{equation}
where $C_{3}$ is given by $C_{3}=2\rho_{*}/Q_{0}=3\rho_{0}/4a_{0}^{3}\lambda^{2}$. Integrating the above differential equation we get the solution as
\begin{equation}
f(Q)=C_{4}+C_{3}\sqrt{Q}+C_{5}Q
\end{equation}
where $C_{4}$ and $C_{5}$ are integration constants. For the GR limit we should have $C_{3}=C_{4}=0$ and $C_{5}=1$.

Using the reconstructed $f(Q)$ in the first FLRW equation (\ref{1frw}) we get
\begin{equation}
\frac{C_{5}}{2}Q-\frac{C_{4}}{2}=\frac{\rho_{*}}{Q_{0}}\left(Q-Q_{0}\right)
\end{equation}
Comparing both sides we get $C_{4}=2\rho_{*}$ and $C_{5}=2\rho_{*}/Q_{0}$. So the fully consistent solution is given by
\begin{equation}
f(Q)=2\rho_{*}+C_{3}\sqrt{Q}+\frac{2\rho_{*}}{Q_{0}}Q
\end{equation}
which can be further written as
\begin{equation}
f(Q)=\frac{2\rho_{0}}{a_{0}^{3}}+\frac{3\rho_{0}}{4a_{0}^{3}\lambda^{2}}\left(\sqrt{Q}+Q\right), ~~~~Q>0    
\end{equation}
Here also we see that, using the first FLRW equation $C_{4}$ and $C_{5}$ has been eliminated, thus reducing the degree of freedom considerably. The above equation may be simplified as
\begin{equation}
f(Q)=E+F(\sqrt{Q}+Q)
\end{equation}
where $E=\frac{2\rho_{0}}{a_{0}^{3}}$ and $F=\frac{3\rho_{0}}{4a_{0}^{3}\lambda^{2}}$.

\begin{figure}[hbt!]
\begin{center}
\includegraphics[height=2.5in]{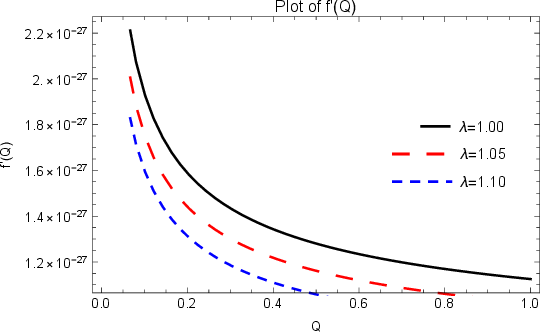}
\caption{Plot of $f'\left(Q\right)$ against $Q$ for different values of $\lambda$ for type-$1$ evolution of scale factor. Other parameters are given by $a_{0}=1$ and $\rho_{0}=10^{-27}$.}
\label{figscalet6}
\end{center}
\end{figure}

\begin{figure}[hbt!]
\begin{center}
\includegraphics[height=2.5in]{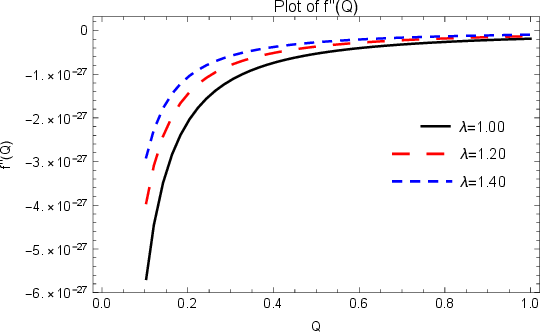}
\caption{Plot of $f''\left(Q\right)$ against $Q$ for different values of $\lambda$ for type-$1$ evolution of scale factor. Other parameters are given by $a_{0}=1$ and $\rho_{0}=10^{-27}$.}
\label{figscalet7}
\end{center}
\end{figure}

For this evolution we have,
\begin{equation}
f'(Q)=F\left(\frac{1}{2\sqrt{Q}}+1\right)
\end{equation}
and
\begin{equation}
f''(Q)=-\frac{F}{4Q\sqrt{Q}}
\end{equation}
We see that $E>0$ and $F>0$. This ensures that $f'(Q)>0$ and $f''(Q)<0$. This is evident in figs.(\ref{figscalet6}) and (\ref{figscalet7}). So we do not get any common parameter region where $f'(Q)>0$ and $f''(Q)>0$ simultaneously hold.

For this type of evolution the second FRW equation becomes,
\begin{equation}
\rho_{eff}+3p_{eff}=\frac{4\lambda^{2}}{3}\left[\coth^{2}{\left(\lambda\left(t-t_{0}\right)\right)-3}\right]
\end{equation}
where $coth$ is the cot hyperbolic function. As $t\rightarrow t_{0}$, we have $\rho_{eff}+3p_{eff}>0$, but as $t\rightarrow \infty$, we get $\rho_{eff}+3p_{eff}<0$. So the energy condition is satisfied in the early times, but it is violated at late times. This shows the presence of exotic matter in the late universe that drives the cosmic acceleration. But in early time the universe is dominated by normal matter, which results in the energy condition being satisfied.
Finally, the non-metricity scalar can be given by
\begin{equation}
Q=6H^{2}=\frac{8}{3}\lambda^{2}\left(1+\frac{a_{0}^{3}}{a^{3}}\right)
\end{equation}
Since the scale factor is real and positive, we should have $Q> \frac{8}{3}\lambda^{2}=Q_{0}$ for this case.

%%%%%%%%%%%%%%%%%%%%%%%%%%%%%%%%%%%%%%%%%%%%%%%%%
\subsection{Type-2 evolution of scale factor}
%%%%%%%%%%%%%%%%%%%%%%%%%%%%%%%%%%%%%%%%%%%%%%%%%
If $AB>0$, then the evolution of scale factor is given by
\begin{equation}
a(t)=a_{0}[\cosh{\left(\lambda\left(t-t_{0}\right)\right)}]^{\frac{2}{3}}
\end{equation}
Unlike Type-I, this branch is non-singular at $\tau=t-t_{0}=0$.
The corresponding equation of state for this type of evolution is given by,
\begin{equation}
\omega_{eff}=\frac{p_{eff}}{\rho_{eff}}=-\left(\coth{\left(\lambda(t-t_{0})\right)}\right)^{2}
\end{equation}
As $t\rightarrow \infty$, $\omega_{eff}\rightarrow -1$, but as $t \rightarrow t_{0}$, $\omega_{eff}$ diverges and tends to a very large number. This shows that at late times we have $\Lambda$CDM-like evolution of the universe, but at early times there is a singularity, with effective EoS diverging to negative infinity. The universe is extremely phantom and probably this indicates a Big Rip during early inflationary phase. This is an unphysical result, since we do not have a matter dominated era in the early time. The second FRW equation takes the form,
\begin{equation}
\rho_{eff}+3p_{eff}=\frac{4\lambda^{2}}{3}\left(\tanh^{2}{\left(\lambda\left(t-t_{0}\right)\right)-3}\right)
\end{equation}
Here $\rho_{eff}+3p_{eff}<0$ for both $t\rightarrow \infty$ and $t\rightarrow t_{0}$. So the energy condition is violated both at early and late times. This is contrary to the observations. The non-metricity scalar for this type of evolution becomes,
\begin{equation}
Q=6H^{2}=\frac{8}{3}\lambda^{2}\left(1-\frac{a_{0}^{3}}{a^{3}}\right)
\end{equation}
The real and positive nature of the scale factor indicates that for this type of evolution we should have $Q<\frac{8}{3}\lambda^{2}=Q_{0}$. 

\begin{figure}[hbt!]
\begin{center}
\includegraphics[height=2.5in]{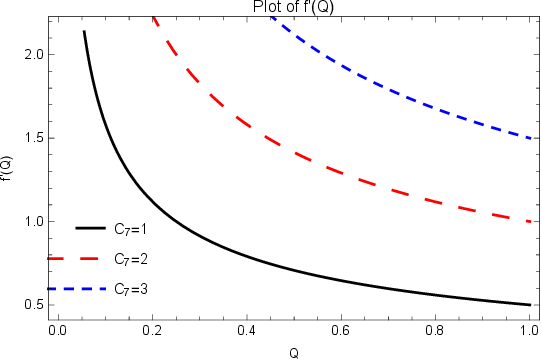}
\caption{Plot of $f'\left(Q\right)$ against $Q$ for different values of $C_{7}$ for type-$2$ evolution of scale factor. Other parameters are given by $a_{0}=1$, and $\lambda=1$, $\rho_{0}=10^{-27}$.}
\label{figscalet8}
\end{center}
\end{figure}

\begin{figure}[hbt!]

\begin{center}
\includegraphics[height=2.5in]{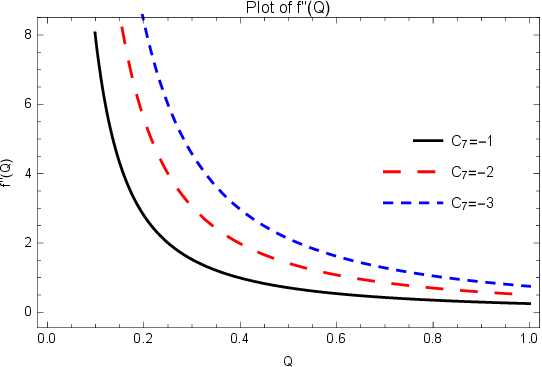}
\caption{Plot of $f''\left(Q\right)$ against $Q$ for different values of $C_{7}$ for type-$2$ evolution of scale factor. Other parameters are given by $a_{0}=1$, and $\lambda=1$, $\rho_{0}=10^{-27}$.}
\label{figscalet9}
\end{center}
\end{figure}

Using eqn.(\ref{rayc}) we get the reconstruction equation for this case as
\begin{equation}
2Qf_{QQ}+f_{Q}=-C_{3}
\end{equation}
where $C_{3}=2\rho_{*}/Q_{0}$. Integrating this we get the solution as
\begin{equation}
f(Q)=C_{6}+C_{7}\sqrt{Q}-C_{3}Q
\end{equation}
For the GR limit we should have $C_{6}=C_{7}=0$ and $C_{3}=-1$.
To constrain the reconstructed model we use the first FLRW equation (\ref{1frw}). Putting the above expression of $f(Q)$ in the first FLRW equation we get
\begin{equation}
-\frac{C_{3}}{2}Q-\frac{C_{6}}{2}=\rho_{*}\left(1-\frac{Q}{Q_{0}}\right)
\end{equation}
Comparing both sides of the above equation we get, $C_{3}=2\rho_{*}/Q_{0}$ and $C_{6}=-2\rho_{*}$. So the fully consistent solution is
\begin{equation}
f(Q)=-2\rho_{*}+C_{7}\sqrt{Q}-\frac{2\rho_{*}}{Q_{0}}Q
\end{equation}
which can be further written as
\begin{equation}
f(Q)=-\frac{2\rho_{0}}{a_{0}^{3}}+C_{7}\sqrt{Q}-\frac{3\rho_{0}}{4a_{0}^{3}\lambda^{2}}Q, ~~~~~~Q>0
\end{equation}
Here we see that there is considerable constraint imposed on the system by the first FLRW equation resulting in loss of degree of freedom. The bove expression can be written as,
\begin{equation}
f(Q)=-E+C_{7}\sqrt{Q}-FQ    
\end{equation}
Using the above expression we get,
\begin{equation}
f'(Q)=\frac{C_{7}}{2\sqrt{Q}}-F
\end{equation}
and
\begin{equation}
f''(Q)=-\frac{C_{7}}{4Q\sqrt{Q}}
\end{equation}
Since $E>0, F>0$, for $C_{7}=0$, $f'(Q)<0$. For $C_{7}>0$ we have $f''(Q)<0$ and for $C_{7}<0$, we have $f'(Q)<0$ and $f''(Q)>0$. In figs.(\ref{figscalet8}) and (\ref{figscalet9}) we have generated plots for $f'(Q)$ and $f''(Q)$ respectively for negative and positive values of $C_{7}$. So here also we see that we do not have a common parametric region where $f'(Q)>0$ and $f''(Q)>0$ holds simultaneously. So, for these models (both type-I and type-II), we have a quasi-stable scenario and we will need a thorough perturbation analysis for a definitive stability conclusion.

The $j=1$ case provides a more physically interesting transitioning expansion history, while making clear that $j=1$ is being used as a theoretically motivated kinematical reconstruction rather than as evidence that the model is observationally identical to $\Lambda$CDM.


%%%%%%%%%%%%%%%%%%%%%%%%%%%%%%%%%%%%%%%%
\section{Discussion and Conclusion}
%%%%%%%%%%%%%%%%%%%%%%%%%%%%%%%%%%%%%%%%
In this study, we have reconstructed the $f(Q)$ gravity model with the Raychaudhuri equation as the primary dynamical component. The method shows that valid modified gravity models within the symmetric teleparallel framework can be found by using the evolution of spacetime congruences. We were able to construct explicit functional forms of $f(Q)$ that accurately mimic the accelerated expansion of the Universe by establishing a relationship between the non-metricity scalar and cosmic kinematics. Two different cases have been discussed in this work. The first one is a straightforward model that accelerates continuously. In line with the examples in the literature, the reconstruction of $f(Q)$ is a straightforward exercise of combination of different powers of $Q$. However, as the powers of $Q$ are not uniquely established, this approach may provide a wide range of models. The GR limit for all the reconstructed models have been investigated.

This approach, which uses the Raychaudhuri equation to theoretically recreate cosmic models, seems to be very effective. In theory, this tool can be used to evaluate a variety of exotic models. By considering a constant deceleration parameter, we have recreated the $f(Q)$ model. The recreated $f(Q)$ model has free parameters, some of which are dependent on the power-law parameter of the expansion scalar. In this case we see that for different values of the parameter $m$ we can realize all the different cosmological eras starting from a early matter dominated phase to a late de-Sitter like evolution. In a second reconstruction effort, we have used a constant jerk parameter. Two different types of scale factor evolution has been considered in this scheme. In type-I we have a matter dominated universe followed by an accelerated expansion. In type-II we have a $\Lambda$CDM-like evolution of the universe in late times, whereas there is a singularity with the effective EoS diverging to negative infinity. The universe is super phantom-like indicating a Big Rip during the early inflationary phase. This is unphysical and hence the type II case is ineffective at early times. The $f(Q)$ theory that has been recovered from the reconstruction is mathematically non-trivial. The functional form of $f(Q)$ contains linear and quadratic terms of the non-metricity scalar. From the models generated from the constant deceleration parameter case, we see that complete background viability can be achieved in the absence of the $\sqrt{Q}$ term. This reduces the model to a power law form depending on the parameter $m$. For the constant jerk parameter reconstructed models, we see that we do not get a common parameter region where we can achieve complete background viability using the derivatives. For these models, we have a quasi-stable scenario and we will need a thorough perturbation analysis for a definitive stability conclusion. Plots are generated for $f'(Q)$ and $f''(Q)$ to check the viability of the models. For different parameter ranges we get different viability scenarios for the constructed model.

The results demonstrate that the Raychaudhuri equation can be used as an inverse reconstruction tool to identify $f(Q)$ functions compatible with prescribed cosmological expansion histories. The reconstructed solutions provide useful analytical examples of the relationship between cosmic kinematics and the non-metricity Lagrangian, while their complete perturbative and local-gravity viability requires further investigation. Cosmological analysis, Constraints from observational datasets, perturbation analysis, stability under scalar and tensor modes, and applications to anisotropic or non-homogeneous cosmic models are possible areas for further research. Deeper understanding of the role of non-metricity in gravitational physics and the late-time dynamics of the universe may be achieved by such investigations.

%%%%%%%%%%%%%%%%%%%%%%%%%%
\section*{Acknowledgments}
%%%%%%%%%%%%%%%%%%%%%%%%%%
PR acknowledges the Inter-University Centre for Astronomy and Astrophysics (IUCAA), Pune, India, for granting a visiting associateship.

%%%%%%%%%%%%%%%%%%%%%%%%%%%%%%%%%%%%%%%
\section*{Data Availability Statement}
%%%%%%%%%%%%%%%%%%%%%%%%%%%%%%%%%%%%%%%
No data was generated or analyzed in this study.

%%%%%%%%%%%%%%%%%%%%%%%%%%%%%%%%%%%%
\section*{Conflict of Interest}
%%%%%%%%%%%%%%%%%%%%%%%%%%%%%%%%%%%%
There are no conflicts of interest.

%%%%%%%%%%%%%%%%%%%%%%%%%%%%%%%
\section*{Funding Statement}
%%%%%%%%%%%%%%%%%%%%%%%%%%%%%%%
There is no funding to report for this article.

%%%%%%%%%%%%%%%%%%%%%%%%%%%%%

\end{document}